\documentclass[a4paper,fleqn]{cas-dc}

\usepackage[numbers]{natbib}
\begin{document}
\let\WriteBookmarks\relax
\def\floatpagepagefraction{1}
\def\textpagefraction{.001}

\shorttitle{Nanopillar Formation}

\shortauthors{Chabowska, Za{\l}uska-Kotur}

\title [mode = title]{ %Growth Pathways and 
Shape Selection in Nanopillar Formation\\
%Emergence of Shape in Nanopillar Growth: A Modeling Approach
}




\author{Marta A. Chabowska}[orcid=0000-0002-8500-3889]

\cormark[1]


\ead{galicka@ifpan.edu.pl}


\credit{Conceptualization, Investigation, Methodology, Validation, Visualization, Writing - original draft}

\author{Magdalena A. Za{\l}uska-Kotur}[orcid=0000-0003-0488-8425]

\cormark[1]

% Footnote of the second author

% Email id of the second author
\ead{zalum@ifpan.edu.pl}


\credit{Conceptualization, Investigation, Methodology, Validation, Writing - review and editing}

\affiliation[1]{organization={Institute of Physics, Polish Academy of Sciences},
            addressline={al. Lotnik\'{o}w 32/46},
            city={Warsaw},
            country={Poland}}

\cortext[1]{Corresponding author}



\begin{abstract}
Crystal growth processes produce a diverse array of surface formations, primarily distinguished by their geometric shapes. While some structures strictly adhere to the underlying crystal symmetry, others exhibit universal circular or oval geometries. Utilizing Vicinal Cellular Automata (VicCA) modeling, we demonstrate that these morphological differences depend on the spatial distribution of the growth potential. Specifically, local potential variations concentrated around surface steps drive the formation of the  lattice symmetry-following structures, whereas global potentials—often originating from defects—generate universal spherical or oval shapes. Furthermore, we illustrate how these morphologies are influenced by the  growth  parameters such  as  sticking coefficient or diffusion  coefficient.
Although the positioning of surface defects is difficult to control, we show that temperature and external particle flux can be effectively used to steer and manipulate surface pattern formation.
\end{abstract}


%\item
%\item
%\item

%\nocite{*}

% Keywords

\maketitle

% Main text
\section{Introduction}{\label{sec:intro}}
The nanoscale morphology of materials, particularly how it develops through the attachment and reorganization of surface deposits originating from atoms or molecules, remains an active and evolving field of research \cite{Nanoscale-Lee,AFM-Zhong,Nanoscale-Atta}. This interest is driven by a wide range of technological applications, including sensors, emitters, lasers, and other optoelectronic devices \cite{quantum-sensors,InGaN-emitter,laser-diode,CdTe-opto}. In surface growth processes, achieving structures with well-defined symmetry, often matching that of the underlying substrate, is essential. Experimentally, morphology and crystallographic orientation can be controlled through various approaches, such as selective area epitaxy \cite{GaN-CGD} vapor-liquid–solid (VLS) growth, and more advanced electric-field-controlled  VLS techniques \cite{electric-VLS}.
Theoretical studies have further demonstrated that the morphology of nanostructures formed during epitaxial growth is strongly governed by surface diffusion dynamics and associated barriers \cite{Rost,Bartelt,Larsson,Krug-book,Chabowska-ACS}.

In this work, we examine how the surface potential energy landscape dictates the resulting patterns of growing nanopillars. We analyze adatom diffusion within this landscape and characterize the specific structures formed on the crystal surface. Furthermore, we investigate how the site-specific sticking probability at kinks and straight steps modulates the final morphology of these nanostructures.

The parameters of the growth model are defined later, in section \ref{sec:model}. The rest of this article is organized as follows. Section \ref{sec:results} is devoted to discussion of the results, which include surface patterns obtained within the surface potential of local and global character. In the last section \ref{sec:conclusion} we summarize our results.

\section{The VicCA Model}{\label{sec:model}}
The surface dynamics are investigated using a (2+1)D VicCA model, which integrates Cellular Automata (CA) and Monte Carlo (MC) techniques to simulate crystal growth \cite{MZK-crystals,Chabowska-ACS,Chabowska-PRB,Chabowska-Vac,Chabowska-JCG}. A defining feature of this model is the decoupling of its core components, allowing for independent parameter control, enhanced computational efficiency, and high adaptability.
The system is conceptualized as two interacting regions: the bulk crystal and a surface layer of diffusing adatoms. While both regions consist of the same atomic species, their dynamics differ significantly; atomic movement within the bulk is constrained by strong interatomic bonds and the presence of the  newly incorporated layers. Growth occurs as adatoms, supplied by an external flux, are adsorbed onto the surface to form a mobile "surface cloud". These atoms diffuse across the lattice until they reach energetically stable sites where they are incorporated into the bulk.

The model employs a hexagonal close-packed) crystal lattice with a hexagonal network of adsorption sites on the (0001) surface. We apply periodic boundary conditions in both lateral directions. To optimize computational efficiency, the model utilizes CA rules for parallel growth updates.
Incorporation into the crystal is site-dependent: step voids possessing four, five, or six nearest-neighbor bonds are filled unconditionally. Kinks with three bonds are occupied according to a specified probability ($p_{k}$), while the attachment probability at a straight step (two free bonds) is scaled by the square of the kink attachment probability $p_s=p_k^2$.In the case of a local potential, as discussed below, the kink attachment probability is defined as $p_k=\rho$, where $\rho$ represents the local adatom density. The scaling of the step attachment probability is implemented by requiring the occupancy of a neighboring site by another particle, resulting in $p_s=\rho^2$. Furthermore, nucleation on a terrace is only permitted if a critical cluster size of five particles is reached.

Surface pattern formation is primarily driven by adatom diffusion, which is governed by the underlying potential energy landscape. In the global potential case, each site is attributed a specific energy value based on its position. For the local potential case, our model incorporates two key energetic features. The first is a series of potential wells ($E_V$) located at the base of each step; these are essential for initiating step meandering. The second is the Ehrlich–Schwoebel  barrier ($E_{ES}$), located at the upper edge of the steps. These barriers hinder downward diffusion but do not affect the upward movement of adatoms.
The presence and magnitude of both features are critical for morphological development. Crucially, this potential landscape is dynamic; the positions of wells and barriers shift in tandem with step displacement, resulting in a strongly coupled evolution of the surface morphology and the energetic field.

\begin{figure*}[pos=htbp]
		\centering
	a)	\includegraphics[width=0.4\textwidth]{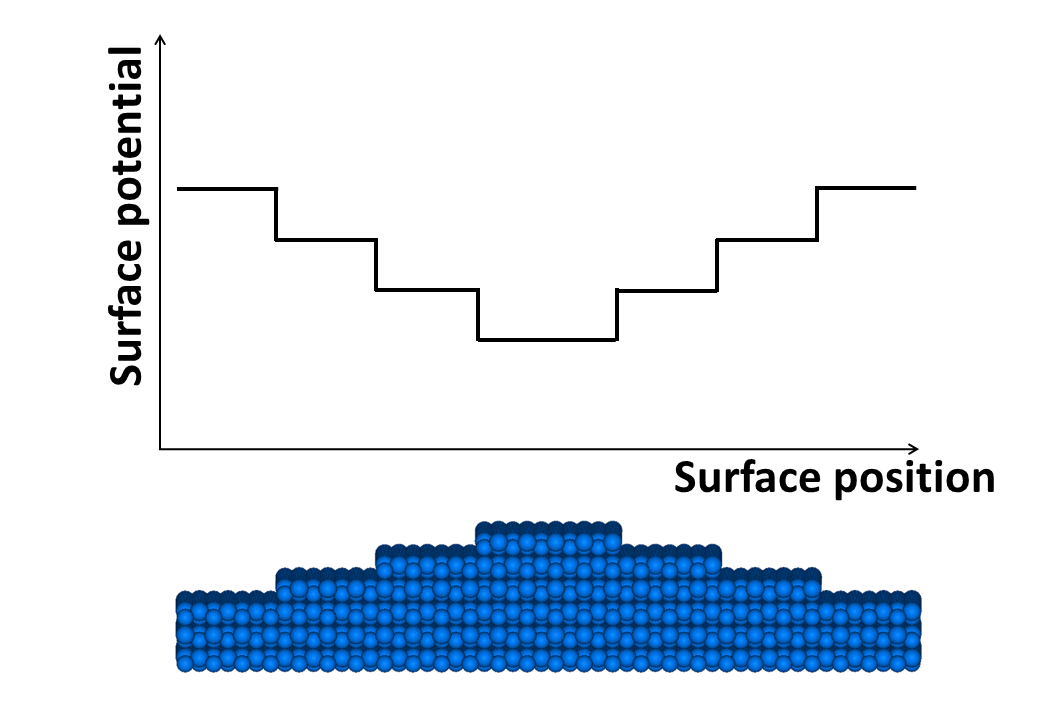}
    b)  \includegraphics[width=0.4\textwidth]{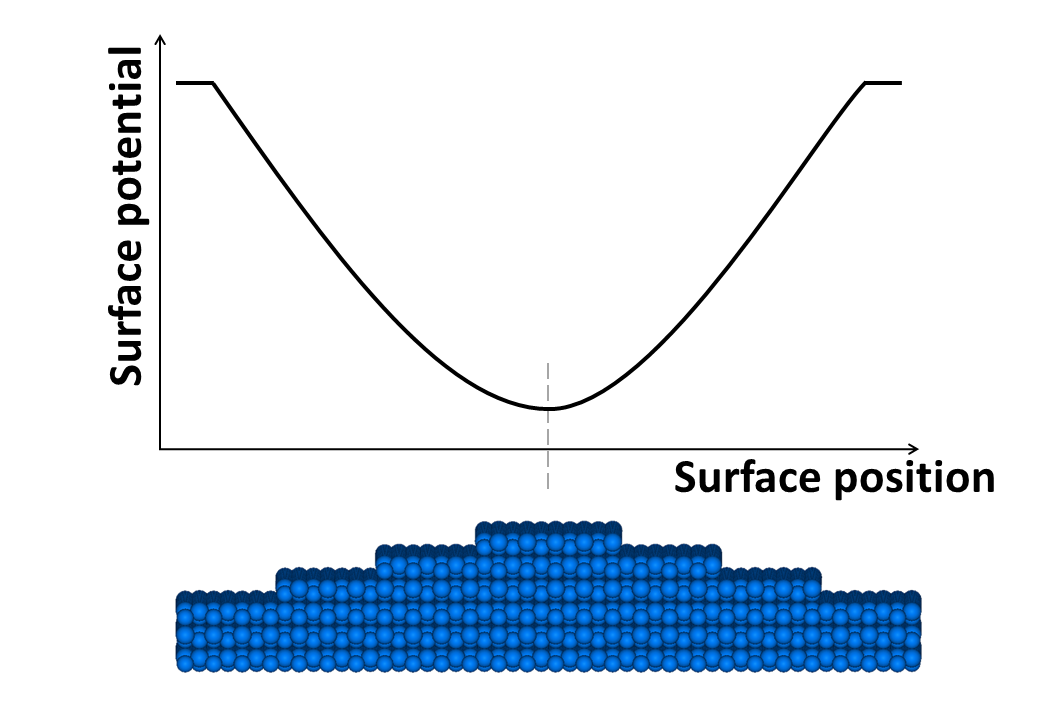}
   	\caption{Visualization of the potential landscape. Side view of the surface with the effective a) local and b) global potential.}
		\label{fig:model}
	\end{figure*}

Diffusion is modeled as a stochastic process where adatoms attempt $n_{DS}$ jumps per time step. The jump probability $P_0$ across a flat terrace is defined by the Boltzmann factor $P_0 = exp(-\beta E_0)$, where $\beta=1/(k_B T)$. Near steps, these probabilities are modulated by the local potentials $E_V$ and $E_{ES}$.
Thus  the  jump  across step is
\begin{equation}
\frac{P_{ES}}{P_0} = \left\{\begin{array}{rcl}
& e^{-(\beta E_V-\beta E_{ES})}& \ \text{out}\\
& 1 & \ \text{in}
\end{array} \right.
\end{equation}
we  use  negative  Shwoebel  barrier, which is subtracted  from  the  depth  of  potential  well for  adatoms that jump  out  of  the  well,  across the  step,  and  it  does  not  change  the  jump down of  the  adatom. Effectively  it  gives potential  that  decreases,  when  particles  go  up (see Fig~\ref{fig:model}).
Such  potential  induces  step  up  flux, allowing  the  transport  of  atoms on  top  of the  built structure.

Unlike standard MC models, where attachment and diffusion are inherently linked to fixed interaction energies, VicCA decouples these mechanisms. This allows us to specify attachment rates and step stiffness independently.
In this approach, a single growth cycle is completed as follows: after $n_{DS}$ diffusion steps, a step attachment event is executed according to the CA rules. Subsequently, the adatom concentration on the surface is replenished to its initial value, $c_0$, by randomly depositing a corresponding number of new adatoms. This sequence concludes a single full simulation step.
Calibration of the physical timescale ($\tau$) and the diffusion-to-flux ratio $D/F$ is managed through the relationship:
$\tau=(\nu n_{DS} e^{-\beta E_0})^{-1}$and $\frac{D}{F} =a^2n_{DS}/c_0$
Temperature influences the system both through explicit jump probabilities and by shifting the effective timescale. Specifically, the number of diffusion steps $n_{DS}$ can be related to temperature changes via $\Delta \beta = \ln(n_{DS}^{-1})/E_0$, where higher $n_{DS}$ values correspond to higher simulation temperatures.

\section{Results and Discussion}{\label{sec:results}}
The primary objective of this paper is to provide a clear and simple explanation of how nanopillars of different shapes are formed depending on the  spatial distribution of the
growth potential. The shapes of these potentials arise from the interactions between atoms, or defects present at the crystal surface. Our investigations were carried out for various growth rates, different depths of the applied surface potential wells, and different heights of the Ehrlich–Schwoebel barrier, when present.

The growth of the structures begins from a seed located at the center of a flat surface. Subsequently, nucleation occurs on the already formed island. The shape of the surface potential favors particles falling into the potential well: the closer a location is to the center of the structure, the deeper the potential well and the more difficult it is for a particle to escape. As a result, the adatom density increases on the tops of the islands, which promotes upward growth.

\subsection{Local potentials}

\begin{figure}[pos=htbp]
		\centering
		\includegraphics[width=0.3\textwidth]{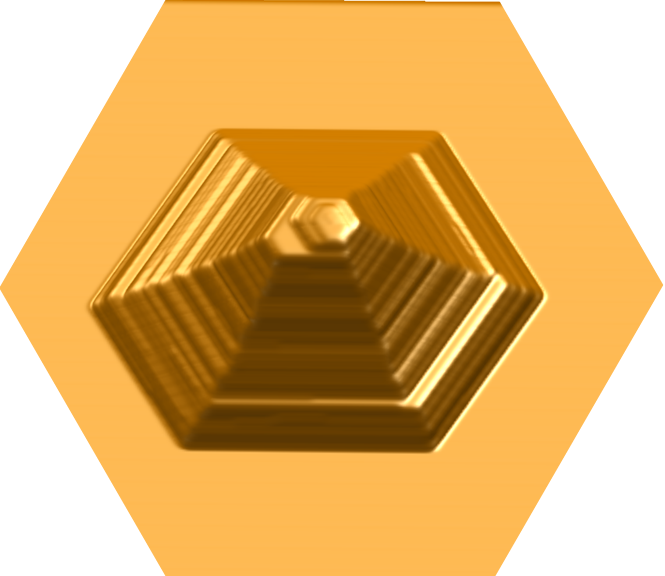}        
   	\caption{A hexagonal nanopillar obtained through adatom diffusion governed by local potential $\beta E_V=4$, $\beta E_{ES}=2$, $n_{DS}=10$, $c_0=0.005$. Number of simulation steps $2^{.}10^6$.}
		\label{fig:results-local}
\end{figure}

Incorporating detailed information regarding local surface interactions and the existing morphology—specifically the positioning of atomic steps—into the definition of the surface potential is both physically justified and intuitively sound. In the initial stage of our study, we introduced a potential landscape that features potential wells of depth $E_V$ at the base of the steps, alongside Ehrlich–Schwoebel  barriers of height $E_{ES}$ at the upper edges. This combined landscape corresponds to the profile illustrated in Figure~\ref{fig:model}a. These localized modifications disrupt the local equilibrium at the step edges, subsequently altering diffusion pathways and inducing a net upward flux of atoms. The results of the simulations conducted using this potential landscape are presented in Figure~\ref{fig:results-local}.

The simulations demonstrate the formation of nanopillar structures exhibiting symmetry consistent with that of the substrate, which in this case is hexagonal. We observe that the potential, which primarily acts at the steps, reveals fine structural details of the analyzed structures, such as the development of well-defined sidewalls. To achieve sustained growth of three-dimensional (3D) structures, the potential well at the bottom of the step must be sufficiently deep, specifically $E_V > 2 k_B T$.  This increases the density of diffusing particles in the vicinity of the steps, as well as the strength of their interaction with the steps, promoting vertical growth.  However, if the well becomes too deep (i.e.,  ~$E_V > 5 k_B T$), the particle density near the steps becomes excessively high. In this regime, the structures grow too rapidly, favoring lateral surface growth over the controlled development of three-dimensional features.

The dimensions of the structures can be controlled by adjusting either the flux of incoming particles or number  of diffusional  steps. Increasing the initial adatom concentration, $c_0$, across the system—which effectively represents a higher external particle flux—results in the growth of larger structures. Our results indicate that also  a prolonged growth phase, characterized by a higher number of diffusion jumps ($n_{DS}$), leads to the formation of nanopillars with both increased height and larger diameters.

\subsection{Defect induced global potentials}

An upward flux of particles may also originate from surface defects. Such defects can significantly modify the potential energy landscape, giving rise to a global potential well that exists independently of local step-related variations (see Figure~\ref{fig:model}b). The profile, spatial extent, and depth of this global potential are determined by the initial surface morphology and the interactions governing adatom dynamics. These interactions, in turn, influence the attachment probabilities at straight steps and kinks. While our model allows these probabilities to be adjusted independently, we assume that since a kink site provides more available bonds than a straight step, the probability of attachment to a straight step is significantly lower.

\begin{figure*}[pos=htbp]
		\centering
	a)	\includegraphics[width=0.3\textwidth]{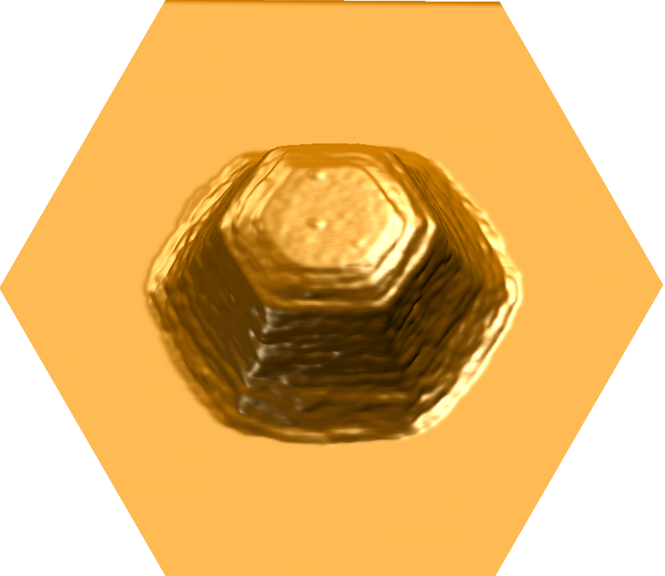}
   	b)  \includegraphics[width=0.3\textwidth]{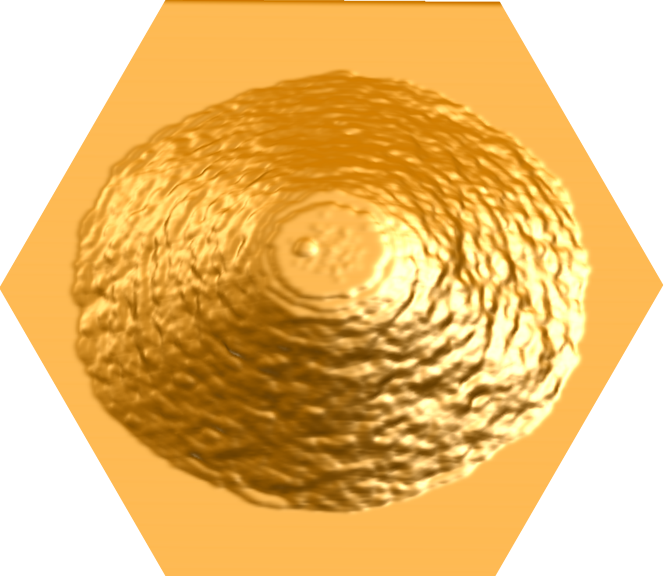}
    c)  \includegraphics[width=0.3\textwidth]{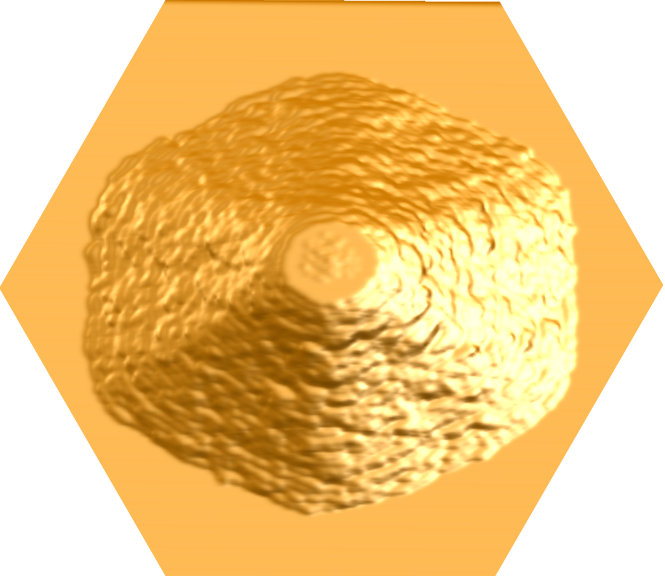}
   	\caption{Nanopillars obtained by adatom diffusion, driven by a cylindrically shaped global potential, with the  probability of an adatom attaching to the kink position is given by  $p_k=e^{-\Delta}$, and  $\Delta=$ a) -3.5 b) -2.2 and c) -1.2. Probability  of  attachment  to  the  step $p_s=p_k^2$,  $n_{DS}=10$, $c_0=0.01$. Number  of simulation steps $2^{.}10^6$.}
		\label{fig:results-global}
\end{figure*}
Figure~\ref{fig:results-global} presents the results of atomic diffusion simulations for the global potential described above. As in the case of local potentials, three-dimensional structures are obtained. However, the final morphology of these structures depends on the particle-attachment probability. A low attachment probability ($P=\exp(-3.5)$) leads to the formation of nanopillars whose symmetry is consistent with that of the substrate - Figure~\ref{fig:results-global}a.  It is worth noting, however, that unlike the previously obtained hexagonal nanopillars shown in Figure 3, the hexagonal edges are less pronounced in this case. In contrast, a higher attachment probability ($P=\exp(-1.2)$) results in nanopillars with hexagonal symmetry at the base and circular symmetry at the top - Figure~\ref{fig:results-global}c. Notably, in this case the orientation of the hexagonal base differs from that observed in the previous case. For intermediate probabilities, a special situation arises in which nanopillars grow with symmetry independent of the substrate - namely, circular symmetry (see Figure~\ref{fig:results-global}b). Structures with such symmetry are difficult to obtain because they occur only within a very narrow range of parameters. 

\begin{figure}[pos=htbp]
		\centering
	\includegraphics[width=0.3\textwidth]{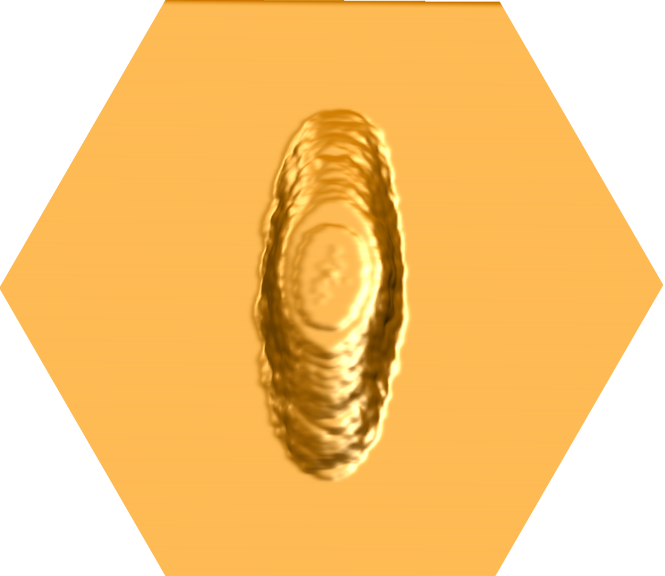}
   	\caption{Nanopillars obtained through adatom diffusion, governed by global potential with an ellipsoidal shape. Probability of an adatom attaching to the kink position is given by  $p_k=e^{-\Delta}$, and  $\Delta=-2.2$. Other parameters are the same as in Fig.~\ref{fig:results-global} }
		\label{fig:results-global-elipse}
\end{figure}

Additional evidence that the final shape of the structures depends strongly on the surface potential is provided by the results obtained for the global ellipsoidal potential shown in Figure~\ref{fig:results-global-elipse}. As can be seen, the asymmetry of the potential becomes the dominant factor governing structure formation, outweighing the influence of substrate symmetry.

\section{Conclusions}{\label{sec:conclusion}}
We have demonstrated that the surface potential energy landscape through which particles diffuse significantly influences the formation and final morphology of nanostructures.  In particular, the dominant origin of the surface potential is crucial in determining the resulting morphology.
When local interactions between particles and surface steps dominate, the potential landscape incorporates local disturbances that modulate diffusion. Key features—such as step edges, the Ehrlich–Schwoebel barriers in their immediate vicinity, and the potential wells at their base—regulate adatom density near the steps. Consequently, sufficiently deep wells facilitate the growth of three-dimensional nanopillars whose symmetry remains consistent with that of the substrate.
In contrast, when surface defects dominate, the potential assumes a more global character, remaining largely independent of local surface fluctuations. In this regime, the sticking probabilities at straight steps and kinks become the decisive factors. Similar to local potentials, low attachment probabilities result in structures with hexagonal symmetry. Conversely, high attachment probabilities lead to 3D structures with a hexagonal base that is rotated by 30° relative to the substrate orientation, with the apex exhibiting circular symmetry.
An interesting regime emerges at intermediate attachment probabilities, where structures with circular symmetry develop regardless of the substrate's underlying symmetry. However, the parameter space for this regime is remarkably narrow; even minor deviations trigger a transition toward hexagonal morphologies.

\section*{Acknowledgments}

The authors express their gratitude to the National Center for Research and Development (grant no. EIG CONCERT-JAPAN/9/56/AtLv-AlGaN/2023). TThey would also like to thank Yoshihiro Kangawa from the RIAM at Kyushu University for their valuable discussions. Most of the calculations were done on HPC facility Nestum (BG161PO003-1.2.05).

 %\appendix
%Appendix A. Supplementary data
%Supplementary data to this article can be found online at 

\printcredits

\bibliographystyle{elsarticle-num}
%\bibliographystyle{cas-model2-names}

%\bibliography{literature}



\end{document}